\documentstyle[preprint,aps,eqsecnum]{revtex}

\tighten

\begin{document}

\draft

\preprint{CRN 96-38}

\title  {
                     Microscopic Study of Superdeformed Rotational
                                  Bands in $^{\bbox{151}}$Tb
        }
\author {
                     N. El Aouad, J.Dobaczewski,\thanks{
                                  Permanent address:
                                  Institute of Theoretical Physics,
                                  Warsaw University, Ho\.za 69,
                                  PL-00681 Warsaw, Poland}
                     J. Dudek, X. Li, W.D. Luo, H. Molique,
                     A. Bouguettoucha, Th. Byrski, F. Beck,
                     and C. Finck
        }
\address{
                     Centre de Recherches Nucl\'eaires,
                     IN$_2$P$_3$--CNRS/Universit\'e Louis Pasteur,
                     F-67037 Strasbourg Cedex 2, France
        }
\author {
                     B. Kharraja
        }
\address{
                     Centre de Recherches Nucl\'eaires,
                     IN$_2$P$_3$--CNRS/Universit\'e Louis Pasteur,
                     F-67037 Strasbourg Cedex 2, France \\
                     Department of Physics, University Chouaib Doukkali,
                     BP20 El-Jadida, Marocco \\
                     University of Notre Dame, Department of Physics, 
                     Notre Dame, IN 46556, USA
        }

\maketitle

\begin{abstract}
Structure of eight superdeformed bands in the nucleus $^{151}$Tb is
analyzed using the results of the Hartree-Fock and Woods-Saxon
cranking approaches.  It is demonstrated that far going
similarities between the two approaches exist and predictions
related to the structure of rotational bands calculated within
the two models are nearly parallel.  An interpretation scenario
for the structure of the superdeformed bands is presented and
predictions related to the exit spins are made.  Small but
systematic discrepancies between experiment and theory, analyzed
in terms of the dynamical moments, ${\cal J}^{(2)}$, are shown to
exist.  The pairing correlations taken into account by using the
particle-number-projection technique are shown to {\it increase}
the disagreement.  Sources of these systematic discrepancies are
discussed -- they are most likely related to the yet not optimal
parametrization of the nuclear interactions used.
\end{abstract}

\pacs{PACS number(s): 21.60.Ev, 21.10.Re, 21.60.Jz, 27.70.+q}

\narrowtext

\section{Introduction}
\label{sec1}

In the present article the results of microscopic calculations
for the high spin properties of the superdeformed (SD) rotational
bands in the nucleus $^{151}$Tb are presented.  This nucleus is a
direct neighbor of the ``doubly magic'' superdeformed nucleus
$^{152}$Dy which is characterized by particularly large gaps in
the neutron and proton deformed single-particle spectra.  We
investigate the rotational bands together with their microscopic
structure in terms of the single-particle orbitals.

The theoretical description of such properties, most often
applied in the literature, is based on the average field
approximation (see Ref.\cite{AFN90a} for a recent review).  Most
of the present knowledge about the nucleonic structure underlying
the properties of the SD rotational bands comes from analyzing
the single-particle Routhians generated by making use of the
rotating Nilsson or deformed Woods-Saxon (WS) Hamiltonians.  The
total nuclear energy as a function of spin is then obtained by
using the Strutinsky total energy expression (see for example
Ref.\cite{WDu95} and references therein).  Similar analyses based
on the Hartree-Fock
(HF)\cite{BON87,BON91,CHB92,ER93,GDB94,DD95,PBF96}, or
relativistic mean field (RMF)\cite{KR89,KR93,AKR96}
approximations are more difficult and have not been, up to now,
applied for global calculations, with the notable exception of
Ref.\cite{AKR96}.  However, they have an advantage of directly
providing the total energy as a function of spin which can be
compared with the experimental data.  In the present work we
apply and compare the WS and HF cranking approximations.

Uncertainties in predictions coming from various models lead to
ambiguities in the interpretation of the experimental results.
The problem of a (non-)uniqueness in the interpretation of the
microscopic structure of SD bands is strongly aggravated by the
fact that the spins of nearly all measured bands in question are
not known.  Moreover, the nuclear interactions responsible for
the single-particle level ordering in the strongly deformed
nuclei are obtained mainly by extrapolating the properties of the
interactions known from spherical and moderately deformed nuclei.

Direct tests of the structure of wave functions of the
single-nucleonic orbitals (via e.g.\ magnetic moments) are
considered extremely difficult from the experimental point of
view due to the very short life-times of the SD high-spin states.
In such a situation we are confronted not only with the problem
of a microscopic interpretation of the quickly growing body of
the experimental information --- but also with the problem of a
cross-check analysis of theoretical predictions coming from
distinct nuclear interactions used.  It will be one of our
objectives to provide such a test by comparing the microscopic
properties calculated within the Strutinsky and HF methods.

{}For a realization of our objective, the experimental knowledge
of eight bands in one single SD nucleus, $^{151}$Tb, is extremely
useful.  At present there are only very few nuclei in which the
number of known SD bands is comparable to that in $^{151}$Tb.
Moreover, the transition energies of those bands behave regularly
as functions of the angular momentum $I$.  Such a regularity
provides the optimal conditions for theoretical verifications of
single-nucleonic configurations via the dynamical moments defined
by
   \begin{equation}\label{eq11}
   {\cal J}^{(2)}=\frac{dI}{d\omega}.
   \end{equation}
This is so, because the abrupt band crossings (relatively
abundant in other nuclei) may otherwise cause strong variations
of ${\cal J}^{(2)}$-moments as a function of $\omega$, thus
rendering the comparisons of the typical behavior (and values) of
the ${\cal J}^{(2)}$-moments with experiment less direct.  In
some cases the band crossings may play an important role in
helping to identify some orbitals via the characteristic crossing
frequencies.  However, we found it preferable to first discuss
the structural elements, such as the global behavior of the
${\cal J}^{(2)}$-moments and their values, in relation with some
characteristic nucleonic configurations.  Another group of
problems related to the band crossings, crossing frequency
systematics and alignments is left for future investigations.

An advantage of basing the verification of the theoretical
results with experiment on the ${\cal J}^{(2)}$-moments is that
this quantity does not depend on the exit spin (the spin
corresponding to the lowest observed transition in an SD band)
while many other quantities do.  For this reason we intend to
focus on the rotational frequency rather than spin as an
independent variable in function of which several microscopically
calculated quantities will be displayed.  However, the question
of the relative spins in the multi-band problem, or equivalently,
of the relations between the exit spins in the bands observed, is
one of the most urgent at present.  For this reason a special
section will be devoted to discussing the total energy vs.  spin
dependence, ${\cal E} = {\cal E}(I)$.

Let us mention here that, similarly to numerous other studies
devoted to the interpretation of the intrinsic configurations of
the SD bands, our work will not be able to give definite answers
in several points.  We feel, nevertheless, that our method allows
to pin down relatively reliably at least the possible alternative
interpretations of the bands in question.

While the lack of knowledge of the exit spins remains one of the
most serious experimental problem, many analyses of the SD nuclei
have benefited from the strong similarities between various bands
in various SD nuclei (the ``identical band'' correlations,
cf.\ Ref.\cite{BHN95} for a recent review).  By comparing the
theoretically predicted structures of the bands related by the
``identicality'' (or ``sameness'') relations, it was possible to
confirm or eliminate certain hypotheses related to the underlying
single-nucleonic structure.  This type of correlations are
abundant in the $^{151}$Tb nucleus and will be used to further
test the microscopic results and predictions.

It is expected that some amount of pairing correlations remains
in the SD bands up to the highest spins measured and that the
pairing correlations decrease with increasing angular momentum.
Although, as we will demonstrate it, in the $^{151}$Tb nucleus
the most important properties of the SD bands such as, e.g., the
behavior of the dynamical moments in function of the rotational
frequency can be described reasonably well without taking into
account the pairing, the systematic small discrepancies which
exist, are increasing when the pairing correlations are taken
into account.  Moreover, the arguments will be given
demonstrating that at least in the case of some parametrizations
of the nuclear interactions another source of small systematic
deviations between theory and experiment may be attributed to
slightly exaggerated nuclear volumes as calculated within the
mean field approaches (cf.\ Sec.\,\ref{sec5} for details).

A number of structural features related to the high-spin behavior
of the SD nuclei in the A$\sim$150 region have been discussed in
the literature.  Early studies were based on the Nilsson or WS
cranking approaches, and performed without pairing\cite{BRA88} or
with the dynamic pairing
correlation\cite{YRS87,JDH88,Naz89,Naz89a} taken into account by
using either the random phase approximation (RPA) or the particle
number projection (PNP) together with the monopole pairing
interaction.  In most of those early studies the experimental
information has been strongly limited, as compared to the amount
of data available today.  In more recent studies, the
WS\cite{SWM94} or the Hartree-Fock-Bogolyubov (HFB)\cite{PBF96}
methods have been applied to a few low-lying bands in a number of
selected A$\sim$150 nuclei with pairing correlations treated in
terms of the Lipkin-Nogami method.

The present study differs from the previously published ones in
the following:  (i) due to a rich experimental information
available now, we can concentrate on systematic discrepancies
between the theoretical and experimental data, (ii) we discuss
the analogies and differences between the self-consistent (HF)
and the non-self-consistent (WS) cranking approaches within the
same analysis (applied simultaneously on a common set of
experimental data), (iii) we analyze simultaneously all available
data in the nucleus in question, (iv) by applying the
particle-number-projected pairing approach we are able to show
that the systematic discrepancies cannot be attributed to the
pairing correlations, and consequently (v) we arrive at a
suggestion that a modification of the underlying interactions
will be needed in order to improve the agreement with data.

\section{Summary of the experimental results}
\label{sec2}

As mentioned earlier, in the $^{151}$Tb nucleus eight SD bands
have been observed.  In addition to the two SD bands discovered
relatively early\cite{PFa89,ThB91}, six new SD bands have been
found and assigned to this nucleus recently\cite{GdF94,BKh95}
using the EUROGAM spectrometer.  Those recent data were obtained
from experiments at the Nuclear Structure Facility, Daresbury
Laboratory, using the $^{130}$Te($^{27}$Al,6n)$^{151}$Tb reaction
at a bombarding energy of 154\,MeV.  The details of the
experimental procedure can be found in\cite{GdF94,BKh95}; here we
only recall that the $\gamma$-rays were detected by the EUROGAM
spectrometer\cite{CWB92,FAB92} which consisted, for this
experiment, of 42 operational Ge detectors.  The data rates in
the individual Ge detectors were 7-9 kHz.  The trigger condition
used to select unsuppressed events was that at least 7 Ge
detectors were involved in a coincidence.  After the suppression
requirement was included, the average number of coincident
detectors dropped to $\sim$\,3.9.  A total data set of 5.5$
\times 10^{8}$ suppressed events with Ge fold $\geq$ 3 was
recorded and after unpacking the high-fold-multiplicity events
led to 3.1 $\times 10^{9}$ $\gamma^4$ coincidences.  The final
spectra were obtained by summing combinations of 3-dimensional
gates set on the SD transitions in a 4-dimensional
analysis\cite{CWB92}.

The assignment of the new 6 excited SD bands to $^{151}$Tb was
based on an unambiguous evidence of known $\gamma$-ray
transitions of the normal deformed nucleus following the
de-excitation from the SD band members.  The SD bands have been
labeled 1,2,3,$\ldots$,8.  The previously reported SD bands have
been labeled 1 (called the yrast band, although its position in
energy with respect to other bands is not experimentally
established) and 2 (the band which shows a remarkable similarity
in terms of the $ \gamma$-ray energies with the SD Band 1 of
$^{152}$Dy).  The other excited SD bands have been arbitrarily
attributed labels ranging from 3 to 8.

Employing the same reaction, the $^{151}$Tb nucleus has been
recently reinvestigated using the Eurogam II array installed at
the Vivitron accelerator of the {\it Centre de Recherches
Nucl\'eaires} in Strasbourg.  This array consisted of 54
high-efficiency Ge detectors (30 tapered coaxial detectors and 24
clover detectors) providing a photopeak efficiency of 7.4\% at
1.33\,MeV.  With the large statistics and high peak-to-background
ratio, it has been possible to obtain a higher precision of the
SD $\gamma$-ray transition energies which in the present study
are used to extract accurate values of the corresponding dynamic
moments.  The experimental results for Bands 1 and 2 are
presented in Table \ref{tab1}.

The intensities (with uncertainties) of the different
bands relative to the yrast band are:
         50(5)\% (Band 2),
         35(5)\% (Band 3),
          6(2)\% (Band 4),
         10(2)\% (Band 5),
          9(2)\% (Band 6),
         11(2)\% (Band 7),
          7(3)\% (Band 8).

Let us also mention that in the determination of the experimental frequencies
and dynamic moments we use the expressions
   \begin{mathletters}\label{eq300}\begin{eqnarray}
     \hbar\omega(I)   &=& \frac{E(I+2)-E(I-2)}{4}
                                       \nonumber      \\
                      &=& \frac{E_\gamma(I+2)+E_\gamma(I)}{4}
                                       \label{eq300a} \\
     {\cal J}^{(2)}(I)&=& \frac{4\hbar^2}{E(I+2)-2E(I)+E(I-2)}
                                       \nonumber      \\
                      &=& \frac{4\hbar^2}{E_\gamma(I+2)-E_\gamma(I)}
                                       \label{eq300b}
   \end{eqnarray}\end{mathletters}\noindent%
where $E_\gamma(I)$=$E(I)$$-$$E(I-2)$.

\section{Woods-Saxon calculations}
\label{sec3}

In this Section we summarize the main points of the calculations
based on the average nuclear field with the deformed WS
potential.  The following three subsections present:  (i) the
structure of the Hamiltonian used, Sec.\,\ref{sec3a}, (ii) some
remarks related to the self-consistency in the case of {\it a
priori} non-self-consistent formalism with the WS potential,
Sec.\,\ref{sec3b}, and (iii) a short description of the actual
variant of the Strutinsky method applied, Sec.\,\ref{sec3c}.  The
details can be found in the literature cited below and will not
be repeated here.  Even though most of the mathematical
expressions are available in the literature, there exist
differences in notation between various authors, and we found
important to recall the definitions of some basic quantities.

\subsection{Deformed Woods-Saxon Hamiltonian and the cranking approximation}
\label{sec3a}

The single-nucleonic energies and wave functions for nucleons moving in the
deformed potential of the WS type,
   \begin{equation}\label{eq31}
         V(\bbox{r};V_{0},\kappa,a,r_{0}) =
         \frac{V_{0} [1 \pm \kappa \frac{N-Z}{N+Z} ]}
         {\{1+\exp [\mbox{dist}_{\Sigma_{0}} (\bbox{r};r_{0} )/a]\}} ,
   \end{equation}
are found by solving the corresponding Schr\"{o}dinger equation.
The
spin-orbit term is included in
the usual deformation-dependent form:
   \begin{eqnarray}\label{eq32}
&&\hat{V}_{\text{SO}}(\bbox{r},\bbox{p};V_{0}^{\text{SO}},\kappa,a_{\text{SO}},
r_{\text{SO}}) =
 \lambda{\left(\frac{\hbar}{2mc}\right)}^{2}
                                                \nonumber \\ &&\times
\left\{\left[ \bbox{\nabla}\frac{V_{0} [1 \pm \kappa \frac{N-Z}{N+Z} ]}
 {\{1+\exp [\mbox{dist}_{\Sigma_{\text{SO}}}
(\bbox{r};r_{\text{SO}} )/a_{\text{SO}}]\}}
 \right]\times{\bbox{p}}\right\}\cdot{\bbox{s}} ,
   \end{eqnarray}
and the Coulomb potential is added for protons.  In the above
relations the ``+'' sign applies for protons and the ``$-$'' sign
for neutrons, while $\bbox{p}$ and $\bbox{s}$ denote the linear
momentum and the spin operators, respectively.  For reasons of
compatibility with previous studies, we keep the factor
${[\hbar/(2mc)]}^{2}$ in the definition of the spin-orbit
potential ($m$ is the nucleonic mass).  Then the spin-orbit
strength parameter, $\lambda$, is a dimensionless quantity.  We
use $\lambda=35$ for neutrons and $\lambda=36$ for protons.

The nuclear surfaces $\Sigma_0$ and $\Sigma_{\text{SO}}$, defining the
central and spin-orbit fields, respectively, are
expressed in terms of the spherical harmonics as
   \begin{equation}\label{eq33}
   \Sigma_{0\text{(SO)}}:\,\,R(\theta,\phi)=
  R_{0\text{(SO)}}c(\alpha)
    \left[1+\sum_{\lambda=0}^{\lambda_{\text{max}}}\alpha_{\lambda\mu}
    Y_{\lambda\mu}(\theta,\phi)\right].
   \end{equation}
In the above relation, $c(\alpha)$ is a function of all the
deformation parameters that guarantees the constant volume for
any deformation, and
\mbox{$R_{0\text{(SO)}}=r_{0\text{(SO)}}A^{1/3}$} with the
central-potential radius parameter $r_{0}=1.347$\,fm for neutrons
and $r_{0}=1.275$\,fm for protons; similarly
$r_{\text{SO}}=1.310$\,fm for neutrons and
$r_{\text{SO}}=1.200$\,fm for protons.  The central-potential
depth-parameters are:  $V_{0}=-49.6$\,MeV and $\kappa=0.86$ while
the diffuseness parameters are $a=0.70$\,fm for the central part
and $a_{\text{SO}}=0.70$\,fm for the spin-obit part of the
potential, independently of the isospin (for details see
Refs.\cite{CJD87} and\cite{DSW82}).  Finally, the geometrical
distance of a point $\bbox{r}$ from the auxiliary surfaces
$\Sigma_{0\text{(SO)}}$ defined above is denoted by
$\mbox{dist}_{\Sigma_{0\text{(SO)}}}$; the latter quantity is
negative for $\bbox{r}$ in the nuclear interior and positive
otherwise.

In our calculations we allow for the total Routhian (for
definition of the Routhians see the following sections)
minimization with the deformation-mesh defined by
   \begin{equation}\label{eq34a}
   \alpha_{20}\,=\,0.400,\,0.425,\,\ldots,0.800\;(17 \;\mbox{points})
   \end{equation}
and
   \begin{equation}\label{eq34b}
   \alpha_{40}\,=\,0.020,\,0.040,\,\ldots,0.180\;(\;9\; \mbox{points})
   \end{equation}
with $\alpha_{22}$ corresponding to 5 ``planes'' for fixed
$\gamma$ deformations [$\alpha_{20}$ =
$\beta$$\cos$($\gamma$) and $\alpha_{22}$ = $\frac{1}{ \sqrt{2}
}$ $\beta$ $\sin$($\gamma$)] as follows
   \begin{equation}\label{eq34c}
\gamma\,=\,-10^{\circ},\-5^{\circ},\,\ldots,10^{\circ}\;(5\; \mbox{points}).
   \end{equation}
We solve the cranking equations of the standard form
   \begin{equation}\label{eq35}
   \hat{h}^{\omega}\psi^{\omega}_{n}=e^{\omega}_{n}\psi^{\omega}_{n}
   \end{equation}
for
   \begin{equation}\label{eq36}
   \mbox{$\hbar \omega$}\,=\,0.00,\,0.05,\,\ldots,1.50\,MeV\;(31\;
   \mbox{points})
   \end{equation}
with
   \begin{equation}\label{eq37}
   \hat{h}^{\omega}\,=\,\hat{T}+V+\hat{V}_{\text{SO}}+V_{\text{C}}
                      \,-\,\omega \hat{I}_{y},
   \end{equation}
where the WS potential $V$ and the related spin-orbit term
$\hat{V}_{\text{SO}}$ are defined in Eqs.\,(\ref{eq31}) and
(\ref{eq32}) while the Coulomb potential $V_{\text{C}}$ is that
of the classical uniform sharp-edge charge distribution
corresponding to the actual nuclear shape.

As always in numerical calculations of the matrix elements used
for constructing the Hamiltonian matrix, both the number of basis
states and the number of points in the Gauss-Hermite integration
formulae have been selected in such a way that the final results
do not depend in any significant way of that choice (for more
details see Ref.\cite{WDu95}).  In our case the number of the
deformed harmonic-oscillator states used in each of the two
parities was about 300.

\subsection{Self-consistency between the potential and density}
\label{sec3b}

In the following, the results of the cranking calculations
performed within the HF approach will be compared with those
obtained for the deformed WS potential.  Although it is not
possible to formulate the self-consistency concept in the two
approaches in the same way, and the WS calculations are {\it a
priori} considered non-self-consistent, it will be of some
importance to recall that our way of defining the deformed WS
potential imposes a correlation between the equipotential
surfaces
   \begin{equation}\label{eq39}
        V(\bbox{r})=\mbox{const.},
   \end{equation}
and the surfaces of the constant density defined by
   \begin{equation}\label{eq39a}
   \rho (\bbox{r} ) = \sum_n{\psi_{n}^{*}(\bbox{r})
                                  \psi_{n}(\bbox{r})}=\mbox{const.},
   \end{equation}
as discussed in detail in\cite{DNO84}. The second-order Thomas-Fermi
expression\cite{JeB75} for the density functional reads
   \begin{eqnarray}\label{eq38}
   \rho(\bbox{r}) &=&
   \frac{1}{3\pi^{2}} {\left[\frac{2m}{{\hbar}^{2}}(\lambda-
   V(\bbox{r}))\right]}^{3/2}
                                \\ &\times&
   \left\{1-\frac{1}{8}\frac{{\hbar}^{2}}{2m}
       \left[\frac{\bbox{\nabla} V(\bbox{r}) \cdot
       \bbox{\nabla} V(\bbox{r})}
       {4{[\lambda-V(\bbox{r})]}^{3}}\right]
       +\frac{\Delta V(\bbox{r})}
       {{[\lambda-V(\bbox{r})]}^{2}} \right\}, \nonumber
   \end{eqnarray}
where the terms of the order ${\cal O}({\hbar}^{4})$ have been neglected.
Requiring that the potential is equal to a constant, cf.\ Eq.\,(\ref{eq39}),
implies by virtue of Eq.\,(\ref{eq31}) that
   \begin{equation}\label{eq301}
        \mbox{dist}(\bbox{r})=\mbox{const.},
   \end{equation}
Using the fact that
   \begin{equation}\label{eq301a}
       \bbox{\nabla} \mbox{dist}(\bbox{r}) \cdot
       \bbox{\nabla} \mbox{dist}(\bbox{r})=1 ,
   \end{equation}
and that
   \begin{equation}\label{eq301b}
       \Delta \mbox{dist}(\bbox{r})=\mbox{const.}
   \end{equation}
for $\bbox{r}$ belonging to the equipotential surface, one obtains
after elementary transformations, cf.\ Ref.\cite{DNO84}, that
also
   \begin{equation}\label{eq302}
       \bbox{\nabla} V(\bbox{r}) \cdot
       \bbox{\nabla} V(\bbox{r})=\mbox{const.}\quad \mbox{and} \quad
       \Delta V(\bbox{r})=\mbox{const.}
   \end{equation}
As a result, at least within the accuracy of ${\cal O}({\hbar}^{4})$
in the Thomas-Fermi approximation, condition (\ref{eq39}) implies
condition (\ref{eq39a}).  Such a correlation between the geometry
of the potential and that of the density is not valid generally.
Neither the Nilsson model nor a deformed WS potential with
another type of argument in the exponent of Eq.\,(\ref{eq31}), as
used sometimes by other authors, gives such a correlation.

Obviously, the spin-orbit term may modify the above result
(cf.\ e.g.\ Ref.\cite{JBB78}), but since this term has, on the average,
matrix elements which are nearly an order of magnitude weaker as
compared to those of the central part, its effect in the
discussed context is also expected to be relatively weak.

Anticipating some of our results presented below, and in
particular surprising similarities between the high-spin behavior
properties obtained by using the self-consistent HF method and a
non-self-consistent approach with the deformed WS Hamiltonian,
one should bear in mind the particular property of definition
(\ref{eq31}) used in this study.

\subsection{Strutinsky method and the total-Routhian calculations}
\label{sec3c}

There are numerous presentations of the Strutinsky method, as
given for example in the original articles\cite{Str67,Str68}, and
then applied to finite potentials (as e.g.\ in Ref.\cite{Bol72}).
Here we closely follow the presentation given in
Ref.\cite{WDu95}, and we only briefly discuss the
total-Routhian-surface calculations, for which several authors
use prescriptions slightly differing in details.

Our calculations of the total nuclear Routhians are based on the expression
   \begin{equation}\label{eq303}
    R^{\omega}=E^{\omega=0}_{\text{macro}}
              +E^{\omega=0}_{\text{micro}}
              +\sum_n{(e^{\omega}_{n}-e^{\omega=0}_{n})} ,
   \end{equation}
where the summation
extends over all the occupied nucleonic states (i.e., in particular takes
into account the particle-hole excitations with respect
to the reference  configuration introduced
in Sec.\,\ref{sec5}). The microscopic
(or shell) energy is defined as usual by
   \begin{equation}\label{eq304}
     E^{\omega=0}_{\text{micro}}=\sum{e^{\omega=0}_{\nu}}-
     {<\sum{e^{\omega=0}_{\nu}}>}_{\text{shell}}
   \end{equation}
where the summation over $\nu$ extends now over all the single-nucleonic states
of the reference configuration. The last term denotes the results of the
Strutinsky averaging procedure. This averaging depends on two
standard
parameters: the smoothing-polynomial order $p$ and the averaging width
$\gamma_{\text{av}}$. In our applications we have $p=6$ and
$\gamma_{\text{av}}=3.2\,\hbar{\omega}_{\text{shell}}$ with
$\hbar{\omega}_{\text{shell}}=41/A^{1/3}$\,MeV. Arguments related to the
stability of the final result with respect to (small) uncertainties in these
parameters can be found in several publications, see e.g.\ Ref.\cite{WDu95}
and references therein.

\section{Hartree-Fock calculations}
\label{sec4}

In the following Sections, some details of a particular
realization of the HF approach in our study will be described;
most of the formal development can be found in the literature and
are also summarized in Ref.\cite{J2D2a}.  Section \ref{sec4a}
presents some technical aspects related to the harmonic
oscillator basis employed, the numerical calculations of the
${\cal J}^{(2)}$-moments are discussed in Sec.\,\ref{sec4b}, and
the use of the quadrupole constraint to stabilize the iteration
process is briefly commented in Sec.\,\ref{sec4c}.

\subsection{Harmonic-oscillator basis}
\label{sec4a}

In the present study we have performed the HF calculations of SD
rotational bands using the Skyrme effective interaction.  {}For
this purpose a new numerical code HFODD has been
constructed\cite{J2D2a} which uses the three-dimensional
Cartesian deformed harmonic oscillator (HO) basis.  The HO basis
has up to now been used in solving the HF equations with the
Gogny interaction\cite{GOG73,GOG75,DEC75,DEC80,BER84}.  On the
other hand, for the Skyrme interaction this method has been used
at early stages of investigations\cite{VBr72} and then abandoned
in favor of the calculations in spatial
coordinates\cite{BON87,BON91}.

The main advantage of using the spatial coordinates consists in
the fact that one may use the same spatial grid of points to
treat nuclei of different deformations, while in the former
method the basis parameters may need to be adjusted with varying
deformation.  (The sensitivity of the results with respect to
this kind of a dependence decreases with an increasing
harmonic-oscillator basis cut-off; in principle the problem
disappears in the limit of an infinite basis).  However, the SD
states have very similar deformations in many neighboring nuclei,
and in such a case this particular deficiency disappears.  We may
then profit from the fact that the calculations which use the HO
basis are in general much more stable, have better convergence
properties, and hence require less computational effort.

The details concerning the HFODD code will be presented
elsewhere\cite{J2D2a}; here we only give a few of its basic
characteristics pertaining to the present application.  The
calculations have been performed using a fixed basis given by the
HO frequencies $\hbar\omega_{\perp} $=11.200\,MeV and
$\hbar\omega_{\parallel}$=6.246\,MeV in the directions
perpendicular and parallel to the harmonic-oscillator
symmetry-axis, respectively.  These values have been obtained by
standard prescriptions developed for diagonalizing the deformed
WS Hamiltonian\cite{CJD87} in the HO basis, and correspond to the
WS potential with the deformations $\beta_2$=0.61 and
$\beta_4$=0.10.  The basis has been restricted to the fixed
number $M$ of HO states, i.e., keeping the lowest $M$
single-particle energies
$\epsilon_{HO}$=$(n_x$+$n_y$+$1)\hbar\omega_{\perp} +
(n_z$+$\frac{1}{2}$)$\hbar\omega_{\parallel}$.  The actual
calculations have been performed with $M$=306.  This corresponds
to the maximal numbers of oscillator quanta equal to 8 and 15 in
the perpendicular and parallel directions, respectively.

The stability of results with respect to increasing the size of
the HO basis is the main concern in all analyses using this
technique, and have been studied in detail in Ref.\cite{J2D2a}.
In fact it is well known that the total energies are
very slowly convergent as functions of the basis size.  Indeed, in
the present calculations a similar effect is seen when we
increase the basis from $M$=306 to $M$=604, which corresponds to
reaching the maximal numbers of 11 and 20 HO quanta in the two
main directions.  Then the total energy of the yrast
configuration of $^{151}$Tb at the rotational frequency of
${\hbar \omega}$=0.6\,MeV decreases from $-$1199.322 to
$-$1201.889\,MeV.  A similar mechanism is well known also from
other type of studies, with the Schr\"{o}dinger equation
diagonalized in a given basis (as for instance with the WS or the
folded Yukawa-type potentials).  Fortunately the relative
(excitation) energies become stable for smaller bases than the
absolute values\cite{J2D2a}.

None of the interesting rotational properties depends in a
significant way on an increase of the basis from $M$=306 to
$M$=604.  For example, we obtain the changes of the average
angular momentum from 59.95 to 59.99\,$\hbar$, of the mass
quadrupole moment from 39.82 to 39.90\,b, and of the mass
hexadecapole moment\cite{units} from 4.21 to 4.24\,b$^2$.  The
relative changes as function of ${\hbar \omega}$ are even
smaller.  Most interestingly, the ${\cal J}^{(2)}$-moment changes
at the same time only from 84.71 to
84.84\,${\hbar}^{2}$\,MeV$^{-1}$.  In view of the fact that the
calculations for $M$=604 are a factor of five slower than those
for $M$=306, we consider it reasonable to keep the latter size of
the basis in all calculations presented.

\subsection{Calculations of the dynamical moments}
\label{sec4b}

An important advantage of using the HO basis consists in the fact
that a rather small number of iterations is sufficient to obtain
a high level of convergence.  For example, in the case of the
yrast configuration in $^{151}$Tb at ${\hbar \omega}$=0.6\,MeV,
starting from the rotating WS state and performing 50 iterations
is enough to obtain the total energy accurate up to 0.5\,keV.
For all values of rotational frequency and with the same number
of iterations performed, the relative precision remains
comparable to the one cited above.

This is very important for the calculation of the second moment of inertia
${\cal J}^{(2)}$. This quantity has to be calculated by a numerical
differentiation, either
of the average angular momentum $I$ \cite{angmo},
   \begin{equation}\label{eq102}
         I = \langle\Psi^{\omega}|\hat{I}_y|\Psi^{\omega}\rangle,
   \end{equation}
with respect to the rotational frequency $\omega$, Eq.\,(\ref{eq11}),
or of the total energy ${\cal E}$,
   \begin{equation}\label{eq101}
   {\cal E} = \langle\Psi^{\omega}|\hat H|\Psi^{\omega}\rangle,
   \end{equation}
with respect to $\omega$:
   \begin{equation}\label{Req12}
   {\cal J}^{(2)}=\frac{1}{\omega}\frac{{d\cal E}}{d\omega}
       +\left(\langle\hat Q_{20}\rangle - \bar Q_{20}\right)
        \frac{2C_{2}}{\omega}\frac{d\langle\hat Q_{20}\rangle}
                              {d\omega} ,
   \end{equation}
where
the cranking wave functions $|\Psi^{\omega}\rangle$ are
obtained from the HF variational equation
   \begin{equation}\label{eq103}
   \delta\langle\Psi^{\omega}|\hat H + C_{2}\left(\langle\hat Q_{20}\rangle
                                                       - \bar Q_{20}\right)^2
                                     -\omega\hat{I}_y|\Psi^{\omega}\rangle = 0.
   \end{equation}
In order to obtain accurate values for ${\cal J}^{(2)}$, the
relative errors of $I$ and ${\cal E}$ as functions of $\omega$
should be kept as small as possible.  The difference of values of
${\cal J}^{(2)}$ calculated from Eqs.\,(\ref{eq11}) and
(\ref{Req12}) is a measure of self-consistency of solutions.  In
our example, the previously quoted value of ${\cal
J}^{(2)}$=84.71\,${\hbar}^{2}$\,MeV$^{-1}$ has been obtained from
Eq.\,(\ref{eq11}), while at the same time Eq.\,(\ref{Req12})
gives ${\cal J}^{(2)}$=84.66\,${\hbar}^{2}$\,MeV$^{-1}$.  These
results have been obtained from the first-order finite-difference
approximation of derivatives between ${\hbar \omega}$=0.60 and
0.61\,MeV.  Together with the discussed uncertainties related to
the restricted size of the basis, these values indicate that our
calculated dynamical moments are accurate to about
0.1\,${\hbar}^{2}$\,MeV$^{-1}$, which is generally a higher
precision than that presently available for the experimental
values of ${\cal J}^{(2)}$.

\subsection{Quadrupole constraint}
\label{sec4c}

The SD states are localized in the pocket of the potential energy
surface which is several MeV above the main minimum.  When the HF
equations are solved by the standard iteration
procedure\cite{RiS80}, which we also use in our calculations, the
wave functions may become gradually polluted by components of the
solutions residing principally in the main well, and this may
slow-down or even preclude any convergence to the secondary
minimum.  In order to prevent this kind of effects we have
included in the minimized functional a quadratic constraint on
the quadrupole moment\cite{Flo73}, i.e., we minimize
   \begin{equation}\label{eq105}
    {\cal E}' = {\cal E}
              + C_{2}\left(\langle\hat Q_{20}\rangle - \bar Q_{20}\right)^2
   \end{equation}
with a small stiffness parameter $C_{2}$=0.01\,MeV/b$^2$ and the
constraint value of $\bar Q_{20}$=42\,b.  This is a very weak
constraint which has no visible effect on the self-con\-sis\-tent
values of the quadrupole moment.  It shifts, however, the
normal-deformed-minimum energy sufficiently high, so that it does
not present any undesired competition with the SD one.

The quadrupole constraint is at the origin of the second term in
Eq.\,(\ref{Req12}), which compensates for the fact that the
quadratic constraint adds to the functional a term which depends
on the rotational frequency.  The magnitude of this second term
depends on the variations of the average quadrupole moment
   \begin{equation}\label{eq104}
                  \langle\hat Q_{20}\rangle =
                  \langle\Psi^{\omega}|\hat Q_{20}|\Psi^{\omega}\rangle
   \end{equation}
along the rotational band (see Sec.\,\ref{sec5}) and on the
difference between $\langle\hat Q_{20}\rangle$ and the target
value $\bar Q_{20}$.  In the present calculations this term can
be for some bands as large as 0.5\,${\hbar}^{2}$\,MeV$^{-1}$,
and, therefore, cannot be neglected when comparing the values of
${\cal J}^{(2)}$ calculated from Eqs.\,(\ref{eq11}) and
(\ref{Req12}).

\section{Theoretical results and comparison with experiment}
\label{sec5}

In this section we present the results of the HF calculations of
the multi-band excitation structure of the $^{151}$Tb nucleus.
In several cases the cross-check results based on the deformed
potential of the WS form will also be presented.  We divide this
discussion into several steps:  Secs.\,\ref{sec5a} and
\ref{sec5c} are devoted to the analysis of the rotational bands
in terms of the single-particle Routhians, total energies, and
dynamical moments ${\cal J}^{(2)}$, while the next ones present
problems related to exit spins, shape changes along the
rotational bands, and pairing.

\subsection{Superdeformed yrast band in $^{\bbox{151}}$Tb}
\label{sec5a}

Most probably, only the lowest energy SD bands are
populated in the present-day experiments on the SD nuclei.  More
precisely, the bands lying not too high in the energy scale, at
spins between, roughly, 60\,$\hbar$ and 80\,$\hbar$, can actually
be detected.  Unfortunately not much more can be said about the
population selectivity and no rules for, for instance, the
population probability as a function of the actual configuration
have been established.  Under these circumstances the only way of
performing a systematic theoretical analysis consists in
calculating numerous bands in excess of the number of those
actually measured and in proceeding with a more detailed and
critical selection analysis next.

\subsubsection{Single-nucleonic energy spectra}
\label{sec5a1}

As the first step the single-particle spectra of the deformed WS
cranking Hamiltonian, cf.\ Sec.\,\ref{sec3}, have been calculated
as functions of the cranking frequency for ${\hbar \omega}$
ranging from 0 to 1\,MeV for a few points corresponding to the
typical quadrupole $\alpha_{20}$ $\sim$\,0.58 and the
hexadecapole $\alpha_{40}$ $\sim$\,0.10 deformations known to
represent well, on the average, the SD nuclei in the $A$$\sim$150
mass range.  Knowing those levels, we define the reference
configuration that corresponds to the lowest-level occupation
scheme for possibly largest ${\hbar \omega}$ range.  The symbols
of the particle-hole configurations used below refer to such a
specific reference configuration as the underlying structure.  It
is worth mentioning that often more than one configuration could
quite well serve as the reference and that the actual choice of
such a configuration is a matter of commodity.  In fact, an
inconvenient choice of the reference would merely result in a
more complicated labeling of the excited bands:  a 1-particle
1-hole excitation with respect to a given reference configuration
may become 2-particles 2-holes, 3-particles 3-holes
etc.\ excitation with respect to the inconveniently chosen reference.

In the present analysis we consider
Hamiltonians having two discrete
symmetries: the parity $\hat {\pi }$ and the
$y$-signature $\hat{R}_y$
   \begin{equation}\label{eq40}
          \hat{R}_{y} = \exp(-i {\pi } \hat{I}_{y}),
   \end{equation}
which has the eigenvalues $r$=$+i$ or $r$=$-i$.  The
reference of our choice is defined in terms of four
types of the occupation numbers: $N_{++}$, $N_{+-}$, $N_{-+}$
and $N_{--}$ referring to positive-parity and signature
$r$=$+i$, positive-parity and signature $r$=$-i$, etc.,
respectively.  We have two sets of such numbers for protons and
neutrons each. In $^{151}$Tb the self-consistent solution for the reference
configuration is defined by occupying the single nucleonic orbitals
according to the following scheme
   \begin{eqnarray}
   \label{eq41}
 \mbox{Neutrons:}\;\;  N_{++}\,&=&\,22,\;\;N_{+-}\,=\,22, \nonumber \\
                       N_{-+}\,&=&\,21,\;\;N_{--}\,=\,21,           \\
 \mbox{Protons:}\;\;\, N_{++}\,&=&\,15,\;\;N_{+-}\,=\,16, \nonumber \\
                       N_{-+}\,&=&\,17,\;\;N_{--}\,=\,17. \nonumber
   \end{eqnarray}
In Fig.\,\ref{fig5.1} we show the single-particle Routhians for
protons and neutrons obtained with the SkM*\cite{BQB82}
parametrization of the Skyrme interaction for the reference
configuration given in Eq.\,(\ref{eq41}).
All HF calculations are in the present
study performed using the complete Skyrme functionals,
cf.\ discussion in Ref.\cite{DD95}.

Let us remark that the interpretation of the experimental SD $^{151}$Tb(1)
yrast band in terms of the configuration implied by relations
(\ref{eq41}) corresponds to the occupation of all the
lowest-lying proton Routhians with the last orbital occupied
$\pi$[6,5,1]3/2($r$=$-i$) (at large $\omega$ taking another
label:  $\pi$[6,4,2]5/2($r$=$-i$), cf.\ Fig.\,\ref{fig5.1}).
This reference configuration has the lowest calculated total energy at a fixed
spin, and therefore, from the theoretical point of view,
we can call it the yrast SD configuration in this nucleus.
Configuration assignment (\ref{eq41}) is confirmed by experimentally
observed band similarities between $^{151}$Tb(1) and
$^{150}$Gd(2), whose several transitions correspond as being
nearly identical.  Indeed, this observation can be viewed as a
direct consequence of exciting the last occupied proton orbital,
$\pi$[3,0,1]1/2($r$=$+i$) in $^{150}$Gd(2), into
$\pi$[6,5,1]3/2($r$=$-i$), the last occupied proton level in
$^{151}$Tb(1).  Since the alignment properties and the ${\cal
J}^{(2)}$-contributions originating from the
$\pi$[3,0,1]1/2($r$=$+i$) orbital are known to be negligible,
Refs.\cite{Naz90,Rag93,DD95}, the assignments of both
configurations in the two nuclei can be viewed as confirmed.

In order to establish the uncertainty margin due to various
parametrizations of the interactions that have been used in the
literature, the results analogous to those in Fig.\,\ref{fig5.1}
but for the SIII parametrization of Ref.\cite{Bei75} are
presented in Fig.\,\ref{fig5.2}.  One can see that the
characteristic gap structures for the protons (in the form of the
sequence of gaps at $Z=66$, $Z=64$, and $Z=62$ visible in
Figs.\,\ref{fig5.1} and \ref{fig5.2}) are very similar in both
parametrizations.  These gap structures are defined by very
similar positions of the $\pi$$[6,5,1]3/2$ and $\pi$$[3,0,1]1/2$
orbitals.  Other proton orbitals, lying above and below those
gaps, differ only slightly in the level ordering.  The
differences at $\omega=0$ in the level positions do not exceed
typically $\sim$\,500 keV.

It will be instructive at this point to compare the
single-particle spectra obtained by using the HF method with the
analogous results obtained by making use of the deformed WS
cranking Hamiltonian.  For this purpose the three-dimensional
total-Routhian surfaces for the $^{151}$Tb have been calculated
in function of the quadrupole ($\beta$,$\gamma$) and hexadecapole
($\beta_{4}$) deformations, cf.\ Sec.\,\ref{sec3a}.  The
total-Routhian surfaces have been calculated in function of the
rotational frequency and the minimum-deformation path composed of
sets of ($\beta$,$\gamma$,$\beta_{4}$)-deformation parameters
that vary with increasing $\hbar\omega$ has been found.  The
deformed WS cranking single-particle spectra obtained in this way
are presented in Fig.\,\ref{fig5.3}.

Let us observe by comparing Figs.\,\ref{fig5.1}, \ref{fig5.2},
and \ref{fig5.3}, that the global features of the single-nucleon
spectra obtained by making use of the deformed average field
model with the WS Hamiltonian and those of the HF with the SkM*
and SIII interactions are quite similar.  What is perhaps most
striking is the fact that the WS Hamiltonian and the HF method
with the SIII parametrization give very similar single-nucleonic
levels indeed.  The similarities are particularly well visible in
the neutron spectra in Figs.\,\ref{fig5.1} and \ref{fig5.3}, and
only to a slightly lesser extent in the proton spectra displayed
in the Figures.  This observation can be interpreted in favor of
both techniques that employ the two different average fields thus
increasing our confidence in the implied results of the
microscopic calculations.

\subsubsection{The ${\cal J}^{(2)}$-moments}
\label{sec5a2}

Since the exit spins of most SD bands are not known, at present
the only possible way of testing the calculated ${\cal E}$ versus
$I$ relation by experiment consists in comparing the resulting
${\cal J}^{(2)}$-moments.  Before proceeding systematically with
such comparisons some additional comments will be useful.  We
begin by discussing the WS calculations which have been performed
here using the so-called ``universal'' parameters.  The
corresponding parameter set has been introduced in
Ref.\cite{DSW82}, and later tested in many detailed calculations
of the high-spin properties in nuclei.  The corresponding
parametrization of the central part of the WS potential employs
the radius-parameter values of $r_{0}$($\nu$) = $1.347$\,fm and
$r_{0}$($\pi$) = $1.28$\,fm for the neutrons and protons,
respectively.  Historically, these parameter values have been
fitted to improve the quality of the single-particle level
spectra rather than to reproduce well the geometrical properties
of nuclei, and it happens that they exceed considerably the more
commonly accepted ``geometrical'' value of
$r_{0}$\,$\simeq$\,$1.24$\,fm.

One of the direct implications of the above difference is that the kinematical
moments of inertia ${\cal J}^{(1)}$,
   \begin{equation}\label{eq42}
   {\cal J}^{(1)}=\frac{I}{\omega } ,
   \end{equation}
calculated from, e.g., the Fermi-gas model with the geometry
defined by using the above radius-parameter values
systematically exceed the rigid-body moments of inertia
calculated from the classical formula with the standard nuclear
matter density. Within a classical but also within a semiclassical model,
the moments of inertia simply scale with the square of the radius
parameter, $r_{0}^2$. As a result one should expect that the moments
of inertia in various bands, calculated by using the ``universal''
parametrization of the WS Hamiltonian will be, on the average, too large.
We can see that such an expectation, valid for the kinematical moments
will apply directly to the dynamical moments (\ref{eq11}),
   \begin{equation}\label{eq43}
   {\cal J}^{(2)}=\frac{dI}{d\omega }=
            {\cal J}^{(1)}+\omega\frac{d{\cal J}^{(1)}}{d\omega },
   \end{equation}
and a scaling of the ${\cal J}^{(1)}$-moments implies
a similar scaling of the ${\cal J}^{(2)}$ ones.

By scaling the calculated ${\cal J}^{(1)}$ and ${\cal
J}^{(2)}$-moments we should, as a consequence, also scale the
related physical quantities:  rotational frequency, energies and
Routhians.  One can see this after noticing that the scaling of
the ${\cal J}^{(1)}$ moment with a factor $f$, that satisfies
relation (\ref{eq42}), for the quantized angular momenta
$I=0,2,4,\ldots$ (which remain independent of any scaling),
implies the necessity of scaling the rotational (cranking)
frequency by a factor of ${f}^{-1}$.  It then follows that the
energies and Routhians also scale as ${f}^{-1}$, so as to
consistently satisfy the canonical relation $E=R+{\hbar\omega}I$.

The results of the calculations of the ${\cal J}^{(2)}$ moment
for the reference (in this case coinciding with the yrast)
configuration of $^{151}$Tb are given in Fig.\,\ref{fig5.4},
where the WS and HF curves are compared with experimental data.
Top part of the Figure gives results with no scaling applied.
One can see, that the calculated values systematically exceed the
experimental ones.  In the WS case we know that, as mentioned
above, the numerical values of the radius-parameters are too
large and we can clearly consider at least a part of the
discrepancy as related to ``too large a volume of the nucleus''.
Therefore, in the bottom part of Fig.\,\ref{fig5.4} we show the
same ${\cal J}^{(2)}$-moments scaled by the factor $f$=0.9.

In the HF case there is no obvious discrepancy between the
calculated and experimental nuclear radii.  Nevertheless, the
calculated ${\cal J}^{(2)}$-moments as well as quadrupole moments
are systematically slightly too large (see
Refs.\cite{SVJ96,Nis96} for recent measurements of the quadrupole
moments and Ref.\cite{WSD96} for the results of systematic HF
calculations of the quadrupole moments and comparison with
experiment).  Such an overestimation is consistently obtained for
many different Skyrme force parameters\cite{DD95}.  At present,
no explanation of this systematic effect can be found.  For the
sake of the presentation we have therefore decided to use the
same phenomenological scaling factor $f$=0.9 to present the
results of HF calculations.

This scaling procedure, as we have checked by an explicit
comparison with the experimental results, works well for all the
bands in the $^{151}$Tb nucleus simultaneously.  Moreover, we
have checked that by using the same scaling factor $f$=0.9 for
many bands and for several nuclei simultaneously one reproduces
well the experimental ${\cal J}^{(2)}$ moments which otherwise
remain too large.  This conclusion is based on comparisons for
the yrast bands in $^{151}$Dy, $^{152}$Dy, $^{153}$Dy,
$^{149}$Tb, and $^{150}$Tb, and also for a few dozens of excited
SD bands in the mass $A$$\sim$150 region.  Light Gadolinum nuclei
($^{146}$Gd and $^{147}$Gd) have strong alignment effects in
their yrast SD bands and are not well suited for the tests
discussed here.  Slopes of the ${\cal J}^{(2)}$ moments in
$^{148}$Gd, $^{149}$Gd, and $^{150}$Gd are overestimated in the
calculations; for the discussion of the latter nucleus see
Sec\,\ref{sec5b}.  Conversely, not introducing the aforementioned
scaling we find the calculated values systematically too high, by
8 to 10\%.

After applying the scaling procedure according to the discussion
above, (cf.\ Fig.\,\ref{fig5.4} for illustration), the ${\cal
J}^{(2)}$-moments are multiplied by the factor of $f$ and the
angular frequencies by a factor of $f^{-1}$.  Then, a very good
correspondence with the $\omega$ dependence of the experimental
data is worth noticing.  Below we present all results scaled
according to the prescription and parameters defined above.  Let
us mention that no physical conclusion of this paper depends on
using or not using the scaling procedure described above; the scaling
can be viewed as merely a ``graphical'' trick to facilitate a
comparison with the experimental data.  Moreover, we have not
attempted any detailed fit of the scaling factor $f$=0.9 which
can still be varied within about $\pm0.02$ without any obvious
general deterioration of the agreement with data.

\subsection{Excited SD bands in $^{\bbox{151}}$Tb}
\label{sec5c}

The problem of the definition of the yrast configuration
discussed in Sec.\,\ref{sec5a1} did not pose any serious
difficulty.  Firstly, in nuclei with the nucleon numbers in the
vicinity of large SD gaps (as e.g.\ $Z$=$66$ and $N$=$86$) the
yrast configurations corresponding to the occupation of the
lowest-lying levels are relatively easily and unambiguously
determined over a relatively long $\omega$-range since no, or
very few level crossings occur at the Fermi level
(cf.\ Fig.\,\ref{fig5.1}).  Secondly, the experimentally dominating SD
band (i.e., the one with the highest population intensity) is
also easily distinguished.  In the case of the other SD bands
neither of the observations applies.  In experiment several bands
are populated with comparable (usually weak) intensities and
their relative excitation energies are totally unknown.  In
theory, numerous close lying bands remain {\it a priori} the
candidates for a much smaller number of the experimental bands to
be associated with.

Moreover, the uncertainties in the relative positions of the
theoretically calculated levels which sometimes influence the
level ordering may reach 500\,keV or more.  This estimate
corresponds to a typical discrepancy between the results obtained
within commonly used average field models (like WS potential or
Skyrme interaction parametrizations, SkM* and SIII as discussed
above and illustrated in Figs.\,\ref{fig5.1}, \ref{fig5.2} and
\ref{fig5.3}) and may, occasionally, become markedly larger.
Consequently, in analyzing the structure of the nucleus in
question we have applied the following approach:

(i) Our primary goal in this section is to recognize within the applied
formalism the most probable
single-nucleonic (particle-hole) structure that underlies the behavior
of the experimentally found bands.

(ii) We are using, for the sake of {\em qualitative} argumentation,
the calculated ordering of various SD bands: the lower the band
energy at the {\em high-spin limit}, the higher the chance to find
such a band among those populated in experiment.

(iii) We use the experimental ${\cal J}^{(2)}$-moments and
verify the possible single-nucleonic configurations by directly
comparing ${\cal J}^{(2)}$($\omega$) with experiment.

Another important piece of information suitable for identifying
single-particle configurations is the sameness of bands in
different nuclei, i.e., the equality of the corresponding
dynamical moments and specific relations which occur for the
relative alignments, see the review in Ref.\cite{BHN95}.  In the
present work we mainly use and analyze the properties of the
dynamical moments, because those of the relative alignments are
very sensitive to the details of the effective
interactions\cite{DD95}.  A readjustment of the interaction
parameters, especially those related to time-odd components of
the mean field, seems to be necessary for a precise theoretical
determination of the relative alignments.

In order to apply the guidelines (ii) and (iii) above, we first
present the results for the lowest-lying one-particle one-hole
excited bands.  In Figs.\,\ref{fig5.5} and \ref{fig5.6} we show
the total energies for the proton and neutron excitations,
respectively.  Here and in the following we label the
configurations by the corresponding Nilsson labels of hole and
particle states, as well as by the total parity and signature
quantum numbers supplemented by (arbitrary) consecutive number
attached to each configuration.  Since in $^{151}$Tb the size of
the proton shell gap is smaller than that for the neutrons, there
are more low-lying proton than neutron excited bands.  Since we
know that there exist uncertainties in the single-particle level
positions that may influence the relative positions of some bands
in a non-negligible manner, the results in Figs.\,\ref{fig5.5}
and \ref{fig5.6} should be treated as a semi-quantitative guide.
The Figures show 10 proton particle-hole and 12 neutron
particle-hole excited bands, respectively, that appear to be the
lowest at $I\sim80\hbar$, i.e., in the spin range where the
superdeformed bands are populated.  We have placed some of the
curves in the upper parts, and some others in the lower parts of
the Figures trying to illustrate an existence of ``families'' of
bands, each family characterized by some specific features like
slopes or band crossings.  In the case of neutrons, the
separation into two families is particularly well visible.  The
differences in this case (cf.\ also Figs.  \ref{fig5.7} and
\ref{fig5.8}) correspond to the structure differences in the two
characteristic hole-orbitals ($\nu[6,5,1]3/2(r$=$+i)$ or
$\nu[6,5,1]3/2(r$=$-i)$, top parts, and $\nu[7,6,1]3/2(r$=$-i)$
bottom parts).

Figures \ref{fig5.7} and \ref{fig5.8} present the theoretical
predictions for the dynamical moments corresponding to the bands
shown in Figs.  \ref{fig5.5} and \ref{fig5.6}, respectively.  One
can see that the theoretically calculated moments form
``families'' characterized by a similar behavior also in the
proton case.  In particular the bands built on the configurations
of the type $\pi[3,0,1](r$=${\pm}i)$ $\rightarrow$
$\pi[6,6,0](r$=$+i)$ and $\pi[6,5,1]3/2(r$=$-i)$ $\rightarrow$
$\pi[6,6,0](r$=$+i)$ are characterized by similar slopes, in
terms of the ${\cal J}^{(2)}$ moments; these slopes are different
from that of the reference band.  On the neutron side, the
similarities between the bands related to exciting the
$\nu[7,6,1]3/2(r$=$-i)$ orbital into the orbitals
$\nu[5,1,4]9/2(r$=${\pm}i)$, $\nu[5,2,1]3/2(r$=${\pm}i)$, and
$\nu[4,0,2]5/2(r$=${\pm}i)$ deserve noticing.  Distinguishing
among these properties will be particularly useful when
attributing the configurations to the experimental bands since,
as mentioned before, several theoretical configurations provide
similar ${\cal J}^{(2)}$ moment dependence on the rotational
frequency, while their energies are also quite close.

\subsubsection{Bands $^{151}Tb(2)$, $^{151}Tb(3)$, and $^{151}Tb(4)$ }
\label{sec5b1}

We proceed with the interpretation of the experimental results
for Bands 2, 3, and 4.  Although the most likely structures of
these bands have already been discussed before in the
literature\cite{Naz90,BKh95,PBF96,AKR96}, we found it instructive
to illustrate the procedure of attributing the theoretical bands
to the experimental ones, the procedure which will be of more
importance for other bands for which an adequate interpretation
remains less easy.

By eliminating the bands whose ${\cal J}^{(2)}$-moments differ
relatively strongly with respect to the experimental result for
Bands 2-4, we arrive at a comparison presented in
Fig.\,\ref{fig5.9}.  Here and in the following we denote the HF
states by their dominant Nilsson labels at {\em high
frequencies}, cf.\ Fig.\,\ref{fig5.1}.  For example, the above
mentioned $\pi[6,6,0]1/2(r$=$+i)$ Routhian originates at low
frequencies from the $\pi[6,5,1]3/2(r$=$+i)$ state.

Experimental results for Bands 2 and 4 agree rather well with
calculations for configurations based on exciting either of the
signature partners of the $\pi[3,0,1]1/2(r$=${\pm}i)$ orbitals
into the $\pi[6,6,0]1/2(r$=$+i)$ state.  As seen in
Fig.\,\ref{fig5.5}, the configuration based on the
$\pi[3,0,1]1/2(r$=$+i$) Routhian has the energy lower by about
1\,MeV than that based on the corresponding signature partner,
and therefore the former should be retained as the most likely
interpretation of Band 2.  Similarly, the dynamical moment of the
$\pi[6,5,1]3/2(r$=$-i)$ $\rightarrow$ $\pi[6,6,0]1/2(r$=$+i)$
configuration, Fig.\,\ref{fig5.9}, middle part, reproduces very
well the data for Band 3, and this supports the corresponding
interpretation (proposed also in Ref.\cite{BKh95}).  The
suggestion of this configuration is based simultaneously on the
fact that it corresponds to a low-energy excitation, as it is
seen from Fig.\,\ref{fig5.5}, and that it provides a good
agreement of the ${\cal J}^{(2)}$ versus $\omega$ dependence with
the experimental one.

On the other hand, as seen in Figs.\,\ref{fig5.1}, \ref{fig5.2},
and \ref{fig5.3}, all calculations give at high frequencies the
low-lying Routhian $\pi[5,3,0]1/2(r$=$+i)$ that might produce
configurations which are lower in energy,
cf.\ Fig.\,\ref{fig5.5}, bottom part, than those based on excitations
to the $\pi[6,6,0]1/2(r$=$+i)$ state (cf.\ also
Fig.\,\ref{fig5.7}, bottom part).

\subsubsection{Bands $^{151}Tb(5)$ and $^{151}Tb(6)$ }
\label{sec5b4}

The experimental results for bands $^{151}$Tb(5) and
$^{151}$Tb(6) indicate a correlation with the yrast sequence of
the isotope $^{150}$Tb\cite{GdF94}.  This suggests that the
neutron Routhian $\nu$[7,7,0]1/2($r$=$-i$),
cf.\ Figs.\,\ref{fig5.1} and \ref{fig5.3}, is not occupied in the
related configurations.  [Due to the level interactions at large
$\omega$-values the label of this level changes into
$\nu$[7,6,1]3/2($r$=$-i$)].  Indeed, calculations indicate that
several neutron particle-hole excitations give the required
dependence of ${\cal J}^{(2)}$ versus $\omega$, as seen in
Fig.\,\ref{fig5.8}.  These are excitations from the
$\nu$[7,6,1]3/2($r$=$-i$) orbital to either of the
$\nu$[4,0,2]5/2($r$=${\pm}i$),
$\nu$[5,2,1]3/2($r$=${\pm}i$), or
$\nu$[5,1,4]9/2($r$=${\pm}i$) orbitals.
A comparison of the dynamical moments corresponding to the
configuration $\nu$[7,6,1]3/2($r$=$-i$) $\to$
$\nu$[5,2,1]3/2($r$=${\pm}i$) is given in Fig.\,\ref{fig5.10} as
an example, while the excitation energies are presented in
Fig.\,\ref{fig5.6}.  One can see that the energies of the latter
configurations are $\sim$\,(600 to 800) keV higher than those of
the former ones, thus favoring the involvement of the neutron
$\nu$[4,0,2]5/2 and $\nu$[5,2,1]3/2 orbitals as a more likely
explanation.

\subsubsection{Bands $^{151}Tb(7)$ and $^{151}Tb(8)$}
\label{sec5b5}

The experimental results for bands $^{151}$Tb(7) and
$^{151}$Tb(8) indicate a correlation with the yrast band in the
$^{149}$Gd nucleus\cite{GdF94}.  Such a correlation suggests that
the last proton intruder level, $\pi$[6,4,2]5/2($r$=$-i$),
remains unoccupied with the corresponding proton on a low lying
orbital with a small ${\cal J}^{(2)}$-contribution.  In fact, the
low-lying $\pi$[5,3,0]1/2($r$=$+i$) orbital visible, e.g., in
Fig.\,\ref{fig5.1} provides an excellent result,
cf.\ Fig.\,\ref{fig5.11}.  However, since a pair of degenerate bands
is seen in the experiment, one should also consider some other
possibilities.  In fact, several neutron configurations mentioned
in the preceding Section remain equally good candidates, and in
Fig.\,\ref{fig5.11} we also present a comparison with the
experiment of the results obtained for the
$\nu$[7,6,1]3/2($r$=$-i$) $\rightarrow$
$\nu$[4,0,2]5/2($r$=${\pm}i$) excitations.

It is important to observe that the similarity between the ${\cal
J}^{(2)}$-moments in $^{149}$Gd(1), and those in the
$^{151}$Tb(7,8) bands is accompanied by noticeable differences in
both the $Q_{20}$ and $Q_{40}$ moments calculated for the bands
with the configurations mentioned.  For instance the proton
excited configuration carries about 0.5\,b more than the neutron
one:  $Q_{20}[\pi(651)3/2 \rightarrow \pi(530)1/2]$ $>$
$Q_{20}[\nu(761)3/2 \rightarrow \nu(402)5/2]$, as it is seen in
Fig.\,\ref{fig5.11}.

{}For the bands $^{151}$Tb(7) and $^{151}$Tb(8) we thus have
three possible configurations which give almost indistinguishable
dynamical moments.  The importance of the above example lies in
the fact that the configurations for rather different quadrupole moments
and/or deformations may still lead to very similar dynamical moments.
Therefore, a direct association of ``identical'' quadrupole
moments and ``identical'' dynamical moments is not supported by
microscopic calculations.  The bands having identical dynamical
moments may still carry quite different deformations and/or
quadrupole moments.

\subsection{Predictions related to spins in the superdeformed bands of
$^{\bbox{151}}$Tb  nucleus}
\label{sec5d}

The problem of the exit spins in SD bands is one of the most
actual in the present-day studies of the nuclear
superdeformation.  Many questions remain half-answered since the
connections between the SD bands and the rest of the nuclear
decay schemes are not known in the $A$$\sim$150 nuclei from the
experimental point of view.  This makes it impossible to measure,
e.g., the absolute alignments or ${\cal J}^{(1)}$-moments, and to
estimate the degree to which the rapid nuclear collective
rotation resembles that of a rigid body.  Such questions are
being posed also in relation to the pairing behavior at high
spins and in many other contexts.

Our calculations allow, in principle, to predict the absolute
(and thus also the relative) exit spins of the SD bands studied.
In practice, however, there are some deficiencies of (all) the
methods used which impose limitations on the predictive power
related to these particular results of calculations.

{}First of all, in our HF calculations the pairing correlations
have been neglected and the obtained results concerning the
alignment effect may introduce at least small uncertainties.
Although it is often believed that the spins calculated at a
given cranking frequency $\omega$ must be lower when the pairing
correlations are included, the actual tendency will depend on the
position of the Fermi level relative to the occupation of the
large-$j$ shells:  the down-sloping (beginning of the shell) and
the up-sloping (upper part of the shell) orbitals contribute to
the total alignment in opposite ways.  Several aspects related to
pairing will be presented in Sec.\,\ref{sec5b}.

We have analyzed the calculated spin versus frequency plots for
all the bands studied in this article.  The experimental spins of
the last levels seen experimentally are not known.  Therefore, we
can only compare the theoretical and experimental {\em differences} of
spin along the calculated and observed bands.  To this end, we
define the theoretical difference of spin as
   \begin{equation}\label{eq402a}
       \Delta^{\text{th}}{I} = I^{\text{th}}(\omega)
                             - I^{\text{th}}(\omega_\circ) + 2\hbar ,
   \end{equation}
where $\omega_\circ$ is calculated from the two lowest
experimental transitions according to the standard expression
(\ref{eq300a}).  This theoretical difference can be compared to
the experimental difference of spin
   \begin{equation}\label{eq402b}
       \Delta^{\text{exp}}{I} = I^{\text{exp}} - I^{\text{exp}}_\circ +2\hbar
                              = 2n\hbar,
   \end{equation}
(as a function of $\omega$) relative to the (unknown) spin 
$I^{\text{exp}}_\circ$ of the
state fed by the lowest transition seen in experiment. This experimental
difference of spin
simply equals twice the number $n$ of the gamma transition
counted from the bottom of the band.  A shift of 2$\hbar$ is
added to both definitions (\ref{eq402a}) and (\ref{eq402b}) to
account for the fact that the lowest experimental value of the
frequency (\ref{eq300a}) ($\omega_\circ$) is known at
$I^{\text{exp}}_\circ$+2$\hbar$.

{}For all the SD bands in $^{151}$Tb, calculated within the HF
method using the SkM* force and with the scaling factor
$f$=0.9 discussed in Sec.\,\ref{sec5a2}, the differences between
$\Delta^{\text{th}}{I}$ and $\Delta^{\text{exp}}{I}$ do not
exceed $1\,\hbar$, for all the frequencies studied, i.e., from
0.3 to 0.9\,MeV/$\hbar$. The same is also true (up to 0.5\,$\hbar$)
for the yrast
band in $^{152}$Dy. On the other hand, similar comparison
gives a disagreement for the yrast band in $^{150}$Gd,
which indicates a necessity to consider the pairing correlations
in this nucleus, see Sec.\,\ref{sec5b}.

The spins calculated at $\omega_\circ$
give predictions for the exit spins $I_\circ$ in the SD bands,
   \begin{equation}\label{eq403}
       I_\circ = \left[I^{\text{th}}(\omega_\circ)\right] - 2\hbar ,
   \end{equation}
where the brackets $[\,]$ denote the rounding to the nearest integer or
half-integer compatible with the total signature and the parity
of the number of particles.
The obtained values of $I_\circ$ are collected in Table
\ref{tab5} together with the summary of the theoretical HF
configurations used.  In some cases where clearly more than one
configuration may be considered, two of them are included in the
Table to illustrate the possible uncertainty of the
configuration.

The results for the $I_\circ$-values have been given in the Table
as differences with respect to the $I^{(1)}_\circ$=69/2\,$\hbar$
value obtained for Band 1 which is used as the reference.  The
representation in terms of differences is used to stress that the absolute 
value of
$I^{(1)}_\circ$ may be considered less certain (given the
uncertainties of the scaling procedure, neglect of pairing, etc.)
as compared to relative alignments between different bands, which
are expected to be much less model dependent. For example,
the exit spin obtained in $^{152}$Dy(1) from the HF calculations
without pairing is $I_\circ$=26\,$\hbar$ (for the state fed
by the 602.4\,keV transition). This is fully compatible with
the exit spin of 57/2\,$\hbar$ obtained in band $^{151}$Tb(2)
(for the next transition of 646.5\,keV) and giving the relative
alignment of +1/2\,$\hbar$, cf.\ Ref.\ \cite{DD95}.

The effects of dynamical correlations caused by pairing on the
alignment mechanism in SD bands of the neighboring nucleus
$^{152}$Dy have been studied, e.g., in Ref.\cite{JDH88}.
These effects are expected to be small, of the order of a few
$\hbar$ at the most, and to be rather regular functions of the
frequency.  The corresponding alignment corrections were
calculated to be negative for several bands in the $A$$\sim$150 nuclei.
We have repeated the analysis of these correlations for the
even-even neighbors of $^{151}$Tb, i.e., for $^{152}$Dy and
$^{150}$Gd, for which the pairing effects are not influenced by
the blocking mechanism, and are easier to extract in a relatively
pure form (cf.\ Sec.\,\ref{sec5b}).

Without pairing, the same value of $I_\circ$=26\,$\hbar$ is for
$^{152}$Dy(1) obtained both within the HF and WS calculations.
As shown in Sec.\,\ref{sec5b}, the inclusion of pairing in the WS
method brings this value down to 24\,$\hbar$.  Because the
pairing influence increases from $^{152}$Dy towards $^{150}$Gd,
we can very tentatively estimate that the HF absolute exit spins
in Terbium obtained without pairing
overestimate the values with pairing by about
4\,$\hbar$. Hence the inclusion of pairing would give the exit spin of
$I^{(1)}_\circ$=61/2\,$\hbar$ for the state fed by the 768.4\,keV
transition in the $^{151}$Tb(1) band.

It is important to observe that the obtained exit-spin values
vary rather markedly from band to band.  In particular, in $^{151}$Tb(1)
the value of $I_\circ$ is significantly larger (by 6\,$\hbar$)
than that predicted for the first excited band $^{151}$Tb(2).

\subsection{Shape evolution with spin}
\label{sec5e}

The shape evolution for various configurations studied in this
article has been determined by calculating the expectation values
of the quadrupole moments, $Q_{20}$ and $Q_{22}$, and the
hexadecapole moments $Q_{40}$, $Q_{42}$, and
$Q_{44}$\cite{units}.  The calculated axial quadrupole and
hexadecapole moments, $Q_{20}$ and $Q_{40}$, decrease with
increasing rotational frequency for all the bands studied.  The
corresponding results for the most important proton moments, i.e.,
$Q_{20}$ and $Q_{40}$ are presented in
Fig.\,\ref{fig5.12}.  In Fig.\,\ref{fig5.13} we show for each
band the proton quadrupole moments in function of the dynamical
moments.  One can see that these relations are nearly linear,
which indicates that the changes in ${\cal J}^{(2)}$ are
correlated with changes of $Q_{20}$. One can interpret this characteristic
result as a dynamical-moment evolution accompanying the nuclear-shape
polarization due to the rotation. 

It is also worth observing that
the proton excited configurations have noticeably higher
quadrupole moments and that their slopes are slightly higher than
those of the neutron excited bands.

Slight but systematic deviations from the axiality are predicted
with increasing rotational frequency.  It is worth mentioning
that the deviations from axiality for the signature-partner
configurations,
$\pi$[3,0,1]1/2($r$=$+i$) $\rightarrow$ $\pi$[6,6,0]1/2($r$=$+i$) and
$\pi$[3,0,1]1/2($r$=$-i$) $\rightarrow$ $\pi$[6,6,0]1/2($r$=$+i$),
attributed to Bands 2 and 4, respectively, have opposite signs,
although they both remain small.

The axial hexadecapole moments $Q_{40}$ are relatively
diversified in that their variations in function of the
configuration and rotational frequency may reach as much as 25\%.
The more exotic hexadecapole moments with the non-zero
$\mu$-components, i.e., $Q_{42}$ and $Q_{44}$, have also been
calculated.  The interest in those moments has grown recently in
relation with the so-called $\Delta I$=2 staggering observed in
transition energies of some rare-earth SD nuclei.  The staggering
could manifest a presence of those exotic multipoles in the
nuclear shapes and in particular, the non-zero $Q_{44}$-moments
could manifest the nuclear $C_4$-symmetry, as suggested in
Ref.\cite{Fli93} and discussed in\cite{HaM94}.  The effect in
question may be very small, as shown by microscopic
calculations\cite{Fra94,Mag95a,Luo95,Don96,Bou95}.
Experimentally, the staggering phenomenon has not been reported
to exist in $^{151}$Tb nucleus.  Nevertheless, we found it of
interest to estimate the order of magnitude of the relevant
hexadecapole moments.  The calculations indicate that the
$Q_{42}$-moments, although typically two orders of magnitude
smaller than the $Q_{40}$-moments, remain non-zero, while the
$Q_{44}$-moments are still at least by a factor of a few smaller.

\subsection{The moments of inertia in the presence of pairing}
\label{sec5b}

The main purpose of the scaling procedure introduced in
Sec.\,\ref{sec5a2} was, on the one hand, to quantify the
systematic deviations from the experiment of the obtained results
for the ${\cal J}^{(2)}$-moments.  On the other hand this
procedure proves indeed that the obtained discrepancies are
systematic since the same scaling factor has been successfully
used in various nuclei (and for all the bands known in each of
them).  The comparisons were made for several dozens of bands;
presentation of these results was limited so far in this paper to
8 bands in $^{151}$Tb only.  Discrepancies of similar order of
magnitude were also present in calculations of other authors, but
were not discussed in any major detail.  Only as a result of an
extended comparison with the experimental data we were lead to a
suggestion that the discrepancies are systematic and thus that
the underlying theories (or parametrizations of the forces) might
be, and most likely should be improved.

As it is well known, the pairing correlations influence the
moments of inertia in a rather characteristic manner.  Typically
(although not generally), the stronger the pairing correlations,
the smaller the kinematical moments ${\cal J}^{(1)}$.  The
purpose of the present Section is to estimate the effect of
pairing on the discussed moments, thus putting more restriction
on the uncertainties produced by the applied nuclear effective
forces.

A technique used here differs from those of Refs.\cite{YRS87}
and\cite{PBF96} in that we perform the particle number projection
(PNP) of the rotating BCS type wave functions {\em before}
variation (in the former publication the RPA technique within the
Nilsson model has been used while in the latter, an approximate
projection in terms of the Lipkin-Nogami method together with the
HFB approximation).  Here we apply the pairing-self-consistent
Bogolyubov method and the PNP before variation, using the standard monopole
pairing Hamiltonian with the constant average matrix elements,
different for the protons and the neutrons (see below), and the
WS cranking approach.

Calculations\cite{PBF96} of the SD bands in $A$$\sim$150 nuclei
have shown that the inclusion of pairing through the
density-dependent zero-range interactions improves the agreement
with data in $^{150}$Gd, but at the same time the agreement in
$^{152}$Dy is destroyed.  It seems therefore, that at present one
needs some more effort, also in the case of the HF approach, to
clarify the problem of the effective interactions used, and
although here we will not be able to carry through such a complex
project, we would nevertheless like to quantify the difficulty
and prepare the starting point for a more advanced discussion.

Below we are going to concentrate on the two neighbor-isotones of
the $^{151}$Tb$_{86}$ nucleus, i.e., $^{150}$Gd$_{86}$ and
$^{152}$Dy$_{86}$.  We believe that studying the two even-even
nuclei has, in the present context, some important advantages.
As our calculations, as well as those of other authors indicate,
in the isotones in question the main pairing contributions come
from the $N$=86 neutron system.  Since the relative
proton/neutron pairing strengths are uncertain, in studying the
weak pairing limit, the two even-even nuclei mentioned offer less
uncertainties as compared to odd-$A$ ones in which the blocking
in the case of the odd proton may cause an additional difficulty.
In other words:  it is not known at present whether at the weak
pairing limit that we are confronted with at high spins the usual
blocking procedure is a sufficient an approximation.

In order to estimate the influence of the pairing on the
alignment process, we have calculated the kinematical and
dynamical moments in function of rotational frequency at fixed
deformations representing the average equilibrium-deformation
values for the two nuclei in question (for details see captions
to Figs.\,\ref{fig5.14} and \ref{fig5.15}).  We have taken $N$
and $Z$ single-particle states into account when constructing the
Bogolyubov transformation for neutrons and protons, respectively.
When parametrizing the pairing constants (average matrix
elements) we took as a starting point the pairing force strength
parameters from Ref.\cite{DMS80}
   \begin{mathletters}
   \begin{eqnarray}\label{eq44}
   {G}_{n}&=&\frac{1}{A}\left[18.95-0.078(N-Z)\right], \\
                   \label{eq45}
   {G}_{p}&=&\frac{1}{A}\left[17.90+0.176(N-Z)\right],
   \end{eqnarray}
   \end{mathletters}%
where they have been adjusted to experimental data on normal
nuclei by applying the pairing-self-consistent Bogolyubov
formalism without the PNP.  When used within the particle
projection formalism, these values have to be decreased, by about
15\% (several aspects related to the PNP technique can be found
in Ref.\cite{WND85} (solvable model) and, e.g., in
Refs.\cite{Naz89a,JDH88} (cranking with the WS Hamiltonian)).  In
the present application we use the reduction factors of $0.85$
for both the neutrons and the protons, i.e., we apply
$0.85\,{G}_{n}$ and $0.85\,{G}_{p}$ in the monopole
pairing Hamiltonian.  In this respect we in fact apply the
parametrization studied earlier in the normal nuclei thus not
introducing new parameters for those nuclei studied in this
paper.  The results corresponding to our PNP calculations are
presented in Figs.\,\ref{fig5.14} and \ref{fig5.15} for the
$^{152}$Dy and $^{150}$Gd nuclei, respectively.  Let us observe
that the high spin results for the dynamical moments in
$^{150}$Gd agree very well with experiment.  Note that the
scaling factor used here, $f$=0.9, is the same as that applied in
the calculations without pairing presented previously.  In other
words, in this nucleus the calculations without pairing would not
agree with the data.

Altogether, similarly as in Ref.\cite{PBF96} the results with
pairing correspond to an increase in ${\cal J}^{(2)}$ moments as
compared to those without pairing.  At lower spins the crossing
observed in the data through the rise in the ${\cal J}^{(2)}$
moments appears at a similar frequency also in the calculations,
although the theoretical crossing frequency is slightly too
large.  In any case such a crossing is to a far an extent a
matter of individual level positions rather than the collective
pairing phase, and as such is of less importance for the
discussion in this Section.

The results for the dynamical moments in $^{152}$Dy present
similar defect at small frequency limit where the calculation
provide a crossing-like structure at frequencies which are
slightly too high.  Again, for most of the spin values, the
dynamical moments obtained with pairing are larger than those
without pairing.  Consequently, the pairing cannot be a remedy
for the too high moments as calculated (without scaling) both
within the WS and HF methods.

Concerning the high spin behavior of the dynamical moments
(slightly different slopes in theory as compared to the data):
we have checked that a shape evolution (ignored in the
calculations presented in Figs.\,\ref{fig5.14} and \ref{fig5.15})
brings in the effects of broad proton crossings whose onsets are
already visible in the Routhians from Fig.\,\ref{fig5.3}.  These
effects are present in a form of a slight modification of the
${\cal J}^{(2)}$ vs.  $\omega$ dependence in the experimental
frequency range at highest spins only when the proton pairing is
taken into account.  The aforementioned modification is
sufficient to provide the correct asymptotic of the ${\cal
J}^{(2)}$ at high spins in $^{152}$Dy.

Having obtained an overall agreement with the data on the
dynamical moments we have also compared the ${\cal
J}^{(1)}$-moments as well.  The agreement of the kinematical
moments can be considered very good for both nuclei when the
experimental exit-spin values (so far not measured) are assumed
to be equal to those calculated according to the method described
in Sec\,\ref{sec5d}, i.e., $I_\circ$=24 and 32\,$\hbar$ for
$^{152}$Dy and $^{150}$Gd, respectively. Incidentally, without pairing, the
same method gives the value of $I_\circ$=38\,$\hbar$ in $^{150}$Gd,
and an incorrect behavior for the spin difference $\Delta^{\text{th}}{I}$.

Let us note that by taking into account the pairing correlations
a good description of the kinematical moments at the high-spin
limit is obtained for the same values of the scaling factor and
pairing constants common for both nuclei, although the effects of
pairing, especially in the neutrons, are calculated to be
markedly different in the two isotones discussed.  In all
compared situations a scaling factor was necessary in order to
bring the too high no-pairing results on the ${\cal J}^{(1)}$ and
${\cal J}^{(2)}$-moments towards the experimental data simultaneously
for both nuclei studied.

{}For all the three nuclei, i.e., $^{150}$Gd$_{86}$,
$^{151}$Tb$_{86}$, and in $^{152}$Dy$_{86}$, the inertia
contribution coming from the neutrons is expected to dominate
that of the protons by about a factor of two; also the pairing
effects are calculated to have about a factor of two more effect
in the neutron case.  Comparison of the results in the top and
the bottom parts of Figs.\,\ref{fig5.13} and \ref{fig5.14}
indicates that at the highest frequency range the differences
between the proton contributions for the case with pairing as
compared to the no-pairing case amount to not more than about
1\,${\hbar}^{2}$\,MeV$^{-1}$ (about 1\%) in $^{152}$Dy nucleus
and about 0.5\,${\hbar}^{2}$\,MeV$^{-1}$ (about 0.5\%) in the
$^{150}$Gd nucleus.  In the case of the neutrons the
corresponding differences are slightly larger but also very small
and thus the pairing is calculated to influence the kinematical
moments at the highest frequency limit by about $3\%$.

At the low frequency range the pairing correlations are expected
to be systematically more important.  The effect of pairing at
$\hbar\omega\sim 0.4\,MeV$ on the moment of inertia is about
$10\%$ in $^{152}$Dy and about $20\%$ in $^{150}$Gd.  The main
difference between the calculations for $^{152}$Dy and $^{150}$Gd
is the smaller deformation in the case of $^{150}$Gd; the
deformations for most of the bands in the $^{151}$Tb nucleus are
in between those for the discussed even-even isotones.
Consequently we should expect that at the low frequency range in
the $^{151}$Tb nucleus, the pairing correlations should still
amount to $\sim 15 \%$ on the average, those of the protons being
negligible for two reasons:  firstly they are very small in the
two neighboring even-even nuclei and, secondly, the $65th$ odd
proton in the $^{151}$Tb is expected to diminish the proton pairing effect
even more due to the blocking arguments.  This is also why the
results for the Terbium nucleus are expected to lie in between
those for the neighboring even-even nuclei studied here.  The
detailed analysis of this nucleus calls first, however, for a
microscopic justification of the scaling factor introduced in the
present study, and for an improved treatment of pairing in an odd
nucleus.

There exists of course an uncertainty margin in the above
estimates.  The dynamical moments are sensitive to the detailed
variation of the single-nucleonic Routhians with the rotational
frequency and a slow alignment of one single orbital may modify
the ${\cal J}^{(2)}$ moment considerably thus increasing the risk
of estimating the collective pairing effect from the influence of
one level.  Secondly, it is impossible to experimentally separate
the proton and neutron contributions and thus in the calculations
one may obtain in principle similar results by increasing, e.g.,
the proton pairing strength while decreasing that of the
neutrons.  Fortunately, in the two nuclei discussed the protons
contribute significantly less to the total inertia as compared to
the neutrons and, in addition, the neutron pairing effects at
$N=86$ in the Dy and Gd isotones differ considerably.
Consequently the second uncertainty mentioned has a rather small
influence on the final result.

\section{Summary and Conclusions}
\label{sec6}

In this article the two most efficient microscopic approaches in
the high-spin nuclear structure calculations, i.e., those based
on the average nuclear field parametrization in terms of the
deformed WS potential and the self-consistent HF approach
together with the cranking method have been applied.  The results
of the calculations employing these two different methods and
also the results of the calculations with two different
parametrizations of the Skyrme interactions have been compared.

Despite of the fact that the formalisms used differ strongly, the
efforts put earlier (by other authors) in optimizing the
parametrizations of the interactions seem to converge:  the
single-nucleon spectra generated by the HF and the WS
Hamiltonians are very similar.  In particular, we have
demonstrated that at least for the nucleus studied the
self-consistent HF solutions and the WS-Strutinsky energy
minimized results lead to similar super-deformed shell closures.
In protons there is a series of shell gaps corresponding to
$Z$=62, 64 and 66 with a characteristic high level-density areas
closing those gaps from below and from above.  The
$\pi$[3,0,1]1/2 orbital separates the former two gaps and the
$\pi$[6,5,1]3/2 orbital the latter two gaps in both SkM* and SIII
parametrizations of the Skyrme interaction and in the WS case.
Of course in the high-level density areas the relative positions
of the single-particle states differ slightly but practically all
the same Nilsson orbitals are represented there.  In the neutron
spectra the $N$=86 SD gap is cut into two ($N$=86 and $N$=85
gaps) by one of the signature partners of the $\nu$[7,7,0]1/2
orbital.  The neutron spectra manifest two systematic differences
(when comparing HF and WS):  (i) the signature splitting of the
$\nu$[7,7,0]1/2 orbital is much stronger in the HF approach
independently of the force, and (ii) below the big gaps we find,
in the decreasing order, the $\nu$[6,4,2]5/2, $\nu$[6,5,1]1/2,
and $\nu$[4,1,1]1/2 orbitals; in the WS case these N=6 orbitals
are inverted.  The inversion, however, did not cause any
significant problem in interpretations of the spectra in the
$^{151}$Tb nucleus.

Theoretical calculations of the dynamical moments indicate a fair
or good agreement for eight bands known experimentally in
$^{151}$Tb.  This is possible, however, only after a
phenomenological scaling factor, common for all bands, is
introduced.  Generally, for higher-lying SD bands a unique
interpretation of the single-nucleonic structure in terms of the
Nilsson labels does not seem possible (at least at present) for
several reasons.  Firstly, the single-nucleonic Routhian spectra
contain levels of different intrinsic characteristics which
remain nearly degenerate over the long ranges of rotational
frequency.  They lead to numerous configurations with very
similar alignment and dynamical-moment properties and of very
close total energies.  Secondly, the band degeneracy (twinned
band effect) does not provide major help in such cases since
several among excited bands may produce nearly identical band
spectra.

The deformation changes in function of the band-configurations
and rotational frequency are very characteristic:  in all bands
discussed not only the quadrupole ($Q_{20}$) and the hexadecapole
($Q_{40}$) moments decrease with increasing frequency, as
observed also by other authors, but also their variation seems
clearly correlated with that of the dynamical and kinematical
moments.  With progressing alignment small triaxialities
(equivalent to the $\gamma$-deformations of the order of a few
degrees) develop as well.

The possibility of the existence of some special symmetries
(like, e.g., an occurrence of the shapes with the $C_4$-symmetry)
has been studied.  No indication for the $Q_{44}$-moments has
been found (an upper limit is predicted at the level of at most
10$^{-3}$$\times$$Q_{40}$).

The calculations allow to predict the values of the exit spins,
i.e., the spins of the lowest observed SD transitions.  These
predictions are given by the values of spin calculated at the
rotation frequency corresponding to the lowest observed
transition, and rounded to the nearest half-integer value
compatible with the total signature.  The differences between the
exit-spin values in different bands are expected to be more
reliable than the corresponding absolute values.

We found that the systematic discrepancies between
the experimental and calculated results exist in both WS and
HF approaches, even if the pairing correlations are taken into
account. This suggests a need for more effort in determining the optimal
parametrizations of the effective interactions used for a description of
the high-spin properties of the strongly-deformed nuclei.

\acknowledgments

We would like to express our thanks to the {\it Institut du
D\'eveloppement et de Ressources en Informatique Scientifique}
(IDRIS) of CNRS, France, which provided us with the computing
facilities under Project no.~960333.  This research was supported
in part by the Polish Committee for Scientific Research under
Contract No.~2~P03B~034~08, and by the computational grant from
the Interdisciplinary Centre for Mathematical and Computational
Modeling (ICM) of the Warsaw University.

\renewcommand{\topfraction}{0.0}
\renewcommand{\bottomfraction}{0.0}
\renewcommand{\floatpagefraction}{0.0}
\setcounter{topnumber}{20}
\setcounter{bottomnumber}{20}
\setcounter{totalnumber}{20}

\clearpage

\begin{table}
\caption[T1]{Experimental $\gamma$-ray energies (in keV) in SD Bands 1 and 2
             in $^{151}$Tb.}
\label{tab1}
\[\begin{array}{ll}
\hline
 ~\mbox{Band 1}~~~~~~~~ &    ~\mbox{Band 2}  \\
\hline
 ~~~~~~.                &    ~646.5(2)       \\
 ~~~~~~.                &    ~691.6(2)       \\
 ~~~~~~.                &    ~737.0(2)       \\
 ~768.4(2)              &    ~782.5(2)       \\
 ~810.9(2)              &    ~828.6(2)       \\
 ~854.0(2)              &    ~874.5(2)       \\
 ~897.5(2)              &    ~921.4(2)       \\
 ~942.7(2)              &    ~968.6(2)       \\
 ~988.4(2)              &    1015.6(2)       \\
 1034.7(2)              &    1063.0(2)       \\
 1081.9(2)              &    1110.9(2)       \\
 1129.9(2)              &    1158.9(2)       \\
 1178.7(2)              &    1206.9(2)       \\
 1228.0(2)              &    1255.0(2)       \\
 1278.0(2)              &    1303.1(2)       \\
 1328.5(2)              &    1351.4(2)       \\
 1379.9(2)              &    1399.7(2)       \\
 1431.4(2)              &    1448.3(2)       \\
 1483.6(2)              &    1496.5(6)       \\
 1535.8(3)              &    1545.2(6)       \\
 1589.6(1.7)            &    ~~~~~~.         \\
\hline
\end{array}\]

\end{table}

\mediumtext
\begin{table}
\caption[T5]{Theoretical configurations of eight bands in $^{151}$Tb
             and the values of exit spins predicted within the HF
             method with the SkM* force and no pairing. The exit-spin
             values correspond to states fed by the last measured
             $\gamma$ transition identified
             in the last column. The value calculated
             for Band 1 equals to $I^{(1)}_\circ$=69/2$\hbar$.
             An estimated effect of pairing (see text) would bring this 
             value down to 61/2$\hbar$.
             Exit spins of all other bands are given with respect to
             $I^{(1)}_\circ$. When more than one configuration could
             be a candidate (e.g.\ in the case of Bands 5 to 7) at most two
             of them are presented in the Table. 
             Spin differences are given in $\hbar$ and the transition
             energies in keV. The second column gives
             the total parities and signatures of the bands analyzed.
             }
\label{tab5}
\[\begin{array}{ccccc}
\hline
\mbox{Band} & \mbox{($\pi$,r)} & \mbox{Configuration} 
            & ~~~~I_\circ-I^{(1)}_\circ
            & ~~E_\gamma^{I_\circ+2\rightarrow I_\circ}  \\
\hline
     1  & ~(+,-i)~ &~ \nu(22,22,21,21), \pi(15,16,17,17)       & ~~0 & 768.4 \\
     2  & ~(-,-i)~ & ~\pi[3,0,1]1/2(r\mbox{=$+i$})\rightarrow
                      \pi[6,6,0]1/2(r\mbox{=$+i$})~            &  -6 & 646.5 \\
     3  & ~(+,+i)~ & ~\pi[6,5,1]3/2(r\mbox{=$-i$})\rightarrow
                      \pi[6,6,0]1/2(r\mbox{=$+i$})~            &  -3 & 727.1 \\
     4  & ~(-,+i)~ & ~\pi[3,0,1]1/2(r\mbox{=$-i$})\rightarrow
                      \pi[6,6,0]1/2(r\mbox{=$+i$})~            &  +3 & 865.8 \\
     5  & ~(+,-i)~ & ~\nu[7,6,1]3/2(r\mbox{=$-i$})\rightarrow
                      \nu[5,2,1]3/2(r\mbox{=$-i$})~            &  -2 & 710.1 \\
     5  & ~(-,-i)~ & ~\nu[7,6,1]3/2(r\mbox{=$-i$})\rightarrow
                      \nu[4,0,2]5/2(r\mbox{=$-i$})~            &  -2 & 710.1 \\
     6  & ~(+,+i)~ & ~\nu[7,6,1]3/2(r\mbox{=$-i$})\rightarrow
                      \nu[5,2,1]3/2(r\mbox{=$+i$})~            &  +3 & 838.2 \\
     6  & ~(-,+i)~ & ~\nu[7,6,1]3/2(r\mbox{=$-i$})\rightarrow
                      \nu[4,0,2]5/2(r\mbox{=$+i$})~            &  +3 & 838.2 \\
     7  & ~(-,+i)~ & ~\pi[6,5,1]3/2(r\mbox{=$-i$})\rightarrow
                      \pi[5,3,0]1/2(r\mbox{=$+i$})~            &  +1 & 753.9 \\
     7  & ~(-,+i)~ & ~\nu[7,6,1]3/2(r\mbox{=$-i$})\rightarrow
                      \nu[4,0,2]5/2(r\mbox{=$+i$})~            &  -1 & 753.9 \\
     8  & ~(-,-i)~ & ~\nu[7,6,1]3/2(r\mbox{=$-i$})\rightarrow
                      \nu[4,0,2]5/2(r\mbox{=$-i$})~            & ~~0 & 785.0 \\
\hline
\end{array}\]
\end{table}
\narrowtext

\clearpage

\begin{figure}[ht]
\caption[fig5.1]{%
Single-particle Routhians calculated for the $^{151}$Tb nucleus
by using the HF cranking formalism with the SkM* interaction.
The solutions correspond to the reference configuration defined
in Eq.\,(\protect\ref{eq41}).  There are four families of curves
corresponding to different parity-signature combinations, namely,
the positive-parity Routhians are denoted with the full
($r$=$-i$) and dash-dotted ($r$=$+i$) lines, and the
negative-parity Routhians are denoted with the dotted ($r$=$+i$)
and long-dashed ($r$=$-i$) lines.  Recall that the positive
(negative) signature $r$=$+i$ ($r$=$-i$) corresponds to the
negative (positive) signature exponent $\alpha=-\frac{1}{2}$
($\alpha=+\frac{1}{2}$), the latter notation used by some other
authors.  The curves are labeled using the standard Nilsson
labels $[N,n_{z},$$\Lambda$$]$$\Omega$ with the sign of the
signature $r$ added at the end.  They correspond to dominant
components of the calculated wave functions at $\omega$$=0$ (left
set of labels) and $\omega$$=0.8$ (right set of labels).
}
\label{fig5.1}
\end{figure}

\begin{figure}[ht]
\caption[fig5.2]{%
Similar to that in Fig.\,\protect\ref{fig5.1} but using the
SIII parametrization of the Skyrme interactions. The reference
configuration has been defined in Eq.\,(\protect\ref{eq41}).
}
\label{fig5.2}
\end{figure}

\begin{figure}[ht]
\caption[fig5.3]{%
Single-particle Routhians analogous to those shown in
Fig.\,\protect\ref{fig5.1} but for the WS cranking Hamiltonian.
With increasing values of $\omega$, the deformation parameters
($\beta_{2}$,$\gamma$,$\beta_{4}$) vary so as to minimize the
total energy (calculated by making use of the Strutinsky method)
of the reference configuration (\protect\ref{eq41}).
}
\label{fig5.3}
\end{figure}

\begin{figure}[ht]
\caption[fig5.4]{%
Comparison of the calculated dynamical moments (in $\hbar^2$/MeV)
for the yrast
configuration in $^{151}$Tb without (top, $f$=1) and with (bottom,
$f$=0.9)
the scaling procedure described in the text.
}
\label{fig5.4}
\end{figure}

\begin{figure}[ht]
\caption[fig5.5]{%
Energy as a function of spin for the proton particle-hole
configurations corresponding to the low-lying excitations in
$^{151}$Tb.  To increase the legibility of the Figure some
configurations are presented in the top part and other in the
bottom part.  The labels at the right-hand-side define the
single-nucleonic particle-hole configurations:  the first label
gives the hole state, while the one following the symbol
($\rightarrow$) corresponds to the particle state.  The reference
configuration is given in Eq.\,(\protect\ref{eq41}).  The labels
on the left-hand-side give the total parity and the total
signature of the given configuration.  In addition, the
configurations are numbered in an arbitrary way, and these numbers
are used as subscripts to facilitate the identification of
configurations.  The lines used to denote different
parity-signature combinations are identical to those used in
Fig.\,\protect\ref{fig5.1}.  The scaling has been applied here to
conform with the standard of presentation in
Fig.\,\protect\ref{fig5.4}.
}
\label{fig5.5}
\end{figure}

\begin{figure}[ht]
\caption[fig5.6]{%
Same as in Fig.\,\protect\ref{fig5.5} for the low-lying
neutron configurations.
}
\label{fig5.6}
\end{figure}

\begin{figure}[ht]
\caption[fig5.7]{%
The dynamical moments (in $\hbar^2$/MeV) of the lowest-energy
bands corresponding to proton excitations presented in
Fig.\,\protect\ref{fig5.5}.  A presence of at least two types of
behavior (two ``families'' of bands) deserves noticing; those
presented in the bottom part differ systematically from those in
the top part.  To facilitate the comparison between the curves
the vacuum (reference) band has been repeated in both, i.e., the
top and the bottom parts.
}
\label{fig5.7}
\end{figure}

\begin{figure}[ht]
\caption[fig5.8]{%
Similar to that in Fig.\,\protect\ref{fig5.7} but for the
low-lying neutron configurations.  Here at least three
characteristic forms of behavior are present. The configurations
placed in the bottom part of the Figure give the slopes which are
systematically smaller than those in the top part.  Among the
latter ones the bands demonstrating the crossings in the vicinity
of the $\hbar\omega\sim0.7$\,MeV are clearly visible.
}
\label{fig5.8}
\end{figure}

\begin{figure}[ht]
\caption[fig5.9]{%
A comparison of dynamical moments (in $\hbar^2$/MeV) for selected
bands presented in Figs.\,\protect\ref{fig5.7} and
\protect\ref{fig5.8} with experimental data for Bands 2 (top), 3
(middle), and 4 (bottom).
}
\label{fig5.9}
\end{figure}

\begin{figure}[ht]
\caption[fig5.10]{%
The dynamical moments (in $\hbar^2$/MeV) for the neutron
configuration proposed for Bands 5 (top) and 6 (bottom), compared
to the experimental data.
}
\label{fig5.10}
\end{figure}

\begin{figure}[ht]
\caption[fig5.11]{%
Illustration of the possible proton and neutron configurations as
those describing the structure of Bands 7 (top) and 8 (middle).
The dynamical moments (in $\hbar^2$/MeV) are nearly
indistinguishable in proton and neutron configurations, while the
corresponding proton quadrupole moments (bottom) are markedly
different.
}
\label{fig5.11}
\end{figure}

\begin{figure}[ht]
\caption[fig5.12]{%
Evolution of the proton axial moments $Q_{20}$ and $Q_{40}$ as
functions of the angular momentum.  The configurations discussed
in the text are shown.  To gain space the bands are labeled by
the $(\pi,r)_n$-symbols which coincide with those used in previous
Figures and are also listed in Table \protect\ref{tab5}.
}
\label{fig5.12}
\end{figure}

\begin{figure}[ht]
\caption[fig5.13]{%
Proton quadrupole moments as functions of the dynamical moments
(in $\hbar^2$/MeV) for all eight bands corresponding to these
observed experimentally.  Note that the high values of the
quadrupole moments correspond to the {\it low} spin limit
(cf.\ also Fig.\,\protect\ref{fig5.12}).
}
\label{fig5.13}
\end{figure}

\begin{figure}[ht]
\caption[fig5.14]{%
Comparison with the experimental
data of the calculated kinematical (bottom) and
dynamical (top) moments (in $\hbar^2$/MeV) for the yrast
configuration in $^{152}$Dy. Here the cranking WS method
was applied with deformations $\beta_2=0.61$
and $\beta_4=0.12$.  A small variation of the deformation with
increasing rotation was not taken into account here since we are
interested in this case in the effect of the global mechanisms
(geometrical scaling, pairing force coupling) similar at any
deformation.  Full lines represent the total quantities, i.e.,
sums of the proton and neutron contributions; labels ``t/pair''
(``t/no-p'') correspond to the pairing correlations being
included (neglected) respectively.  Separate proton and neutron
contributions with and without pairing are also shown.  The
scaling factor is $f$=0.9.  The ${\cal J}^{(1)}$ moments are
obtained from the experimental data by applying the calculated
exit-spin value of $I_\circ$=24 for the state fed by the
$E_\gamma$=602.4\,keV transition.
}
\label{fig5.14}
\end{figure}

\begin{figure}[ht]
\caption[fig5.15]{%
Similar as in Fig.\,\protect\ref{fig5.14}, but for the yrast
configuration in $^{150}$Gd.  Deformations used:  $\beta_2=0.58$
and $\beta_4=0.09$.  The ${\cal J}^{(1)}$ moments are obtained
from the experimental data by applying the calculated exit-spin
value of $I_\circ$=32 for the state fed by the
$E_\gamma$=815\,keV transition.
}
\label{fig5.15}
\end{figure}
%
%

\begin{thebibliography}{99}
\bibitem{units} We use the quadrupole and hexadecapole moment operators
                normalized as\\
                $Q_{2\mu}$=$\sqrt{16\pi/5}r^2Y_{2\mu}$
                ($Q_{20}$=$2z^2$$-$$x^2$$-$$y^2$),
                and $Q_{4\mu}$=$r^4Y_{4\mu}$
\bibitem{angmo} For technical reasons explained in
                Ref.\protect\cite{J2D2a} we use the $y$ axis as the
                cranking axis. In the present study we identify
                the projection of the angular momentum on the cranking
                axis with the total angular momentum.
\end{thebibliography}
\end{document}